\documentclass[conference]{IEEEtran}
\IEEEoverridecommandlockouts

\usepackage{amsmath,amssymb,amsfonts}
\usepackage{algorithmic}
\usepackage{graphicx}
\usepackage{textcomp}
\usepackage{xcolor}
\def\BibTeX{{\rm B\kern-.05em{\sc i\kern-.025em b}\kern-.08em
    T\kern-.1667em\lower.7ex\hbox{E}\kern-.125emX}}

\usepackage[acronyms]{glossaries}
    \newacronym{3gpp}{3GPP}{3rd Generation Partnership Project}
    \newacronym{awgn}{AWGN}{additive white Gaussian noise}
    \newacronym[\glslongpluralkey={angles of arrival}]{aoa}{AoA}{angle of arrival}
    \newacronym{bs}{BS}{base station}
    \newacronym[\glslongpluralkey={channel covariance matrices}]{ccm}{CCM}{channel covariance matrix}
    \newacronym{ce}{CE}{channel estimation}
    \newacronym{csi}{CSI}{channel state information}
    \newacronym{cme}{CME}{conditional mean estimator}
    \newacronym{cg}{CG}{conditionally Gaussian}
    \newacronym{dft}{DFT}{discrete Fourier transform}
    \newacronym{dl}{DL}{deep learning}
    \newacronym{elbo}{ELBO}{evidence lower bound}
    \newacronym{idft}{IDFT}{inverse \acrshort{dft}}
    \newacronym{kl}{KL}{Kullback-Leibler}
    \newacronym{lmmse}{LMMSE}{linear \acrshort{mmse}}
    \newacronym{ls}{LS}{least squares}
    \newacronym{mimo}{MIMO}{multiple-input multiple-output}
    \newacronym{ml}{ML}{machine learning}
    \newacronym{mmse}{MMSE}{minimum mean square error}
    \newacronym{mse}{MSE}{mean square error}
    \newacronym{nn}{NN}{neural network}
    \newacronym{nmse}{NMSE}{normalized \acrshort{mse}}
    \newacronym{pc}{PC}{pilot contamination}
    \newacronym{quadriga}{QuaDRiGa}{QUAsi Deterministic RadIo channel GenerAtor}
    \newacronym{simo}{SIMO}{single-input multiple-output}
    \newacronym{snr}{SNR}{signal to noise ratio}
    \newacronym{sinr}{SINR}{signal to interference plus noise ratio}
    \newacronym{tdd}{TDD}{time division duplex}
    \newacronym{ue}{UE}{user equipment}
    \newacronym{uma}{UMa}{Urban Macro-Cell}
    \newacronym{ula}{ULA}{uniform linear array}
    \newacronym{vae}{VAE}{variational autoencoder}
\usepackage[hidelinks]{hyperref}
\usepackage{amssymb}
\usepackage{amsmath}
\usepackage{mathtools}
\usepackage{bm}
    \DeclareMathOperator{\diag}{diag}
    \DeclareMathOperator{\Expectation}{E}
    \DeclareMathOperator{\cond}{\:\!|\:\!}
    \DeclareMathOperator{\dif}{d \!}
    \DeclareMathOperator{\Normal}{\mathcal{N}}
    \DeclareMathOperator{\Uniform}{\mathcal{U}}
    \DeclareMathOperator{\Identity}{\mathbf{I}}
    \newcommand*{\hermitian}{^{\mkern-1.5mu\mathrm{H}}}
    \newcommand*{\transpose}{^{\mkern-1.5mu\mathrm{T}}}
    \DeclarePairedDelimiterX{\infdivx}[2]{(}{)}{%
      #1\;\delimsize\|\;#2%
    }
    \newcommand{\kldiv}{D_{\text{KL}}\infdivx}
    \newcommand{\imaginaryJ}{\mathrm{j}}
\usepackage[numbers,sort&compress]{natbib}
\usepackage{siunitx}
\usepackage[skip=5pt]{caption}

\begin{document}

\title{Addressing Pilot Contamination in Channel Estimation with Variational Autoencoders
    \thanks{This work is funded by the Bavarian Ministry of Economic Affairs, Regional Development, and Energy within the project 6G Future Lab Bavaria. The authors acknowledge the financial support from the Federal Ministry of Education and Research of Germany, project ID: 16KISK002.}
}

\author{\IEEEauthorblockN{Amar Kasibovic, Benedikt Fesl, Michael Baur and Wolfgang Utschick}
\IEEEauthorblockA{TUM School of Computation, Information and Technology\\Technical University of Munich, Munich, Germany\\Email: \{\href{mailto:amar.kasibovic@tum.com}{amar.kasibovic}, \href{mailto: benedikt.fesl@tum.de}{benedikt.fesl}, \href{mailto: mi.baur@tum.de}{mi.baur}, \href{mailto: utschick@tum.de}{utschick}\}@tum.de}}

\maketitle

\begin{abstract}
    \Acrfull{pc} is a well-known problem that affects massive \acrfull{mimo} systems.
    When frequency and pilots are reused between different cells, \acrshort{pc} constitutes one of the main bottlenecks of the system's performance.
    In this paper, we propose a method based on the \acrfull{vae}, capable of reducing the impact of \acrshort{pc}-related interference during \acrfull{ce}.
    We obtain the first and second-order statistics of the \acrfull{cg} channels for both the \acrfullpl{ue} in a cell of interest and those in interfering cells, and we then use these moments to compute conditional linear \acrlong{mmse} estimates.
    We show that the proposed estimator is capable of exploiting the interferers' additional statistical knowledge, outperforming other classical approaches.
    Moreover, we highlight how the achievable performance is tied to the chosen setup, making the setup selection crucial in the study of multi-cell \acrshort{ce}.
\end{abstract}

\begin{IEEEkeywords}
    \acrlong{ce}, generative model, pilot contamination, \acrlong{vae}.
\end{IEEEkeywords}

\begin{figure}[b]
    \onecolumn
    \scriptsize
    This work has been submitted to the IEEE for possible publication. Copyright may be transferred without notice, after which this version may no longer be accessible.
    \vspace{-1.5cm}
    \twocolumn
\end{figure}

\section{Introduction}
Massive \acrshort{mimo} has recently established itself as a key technology in modern cellular networks, and \acrfull{ml} is a promising tool for overcoming some of the biggest challenges brought by massive \acrshort{mimo} \cite{BJORNSON20193}.
One of these obstacles is \acrshort{pc}, which limits performance when the same frequencies and correlated pilots are reused among different cells.
In particular, when a \acrshort{ue} of interest and an unwanted \acrshort{ue} in a different cell share the same pilot, the unwanted \acrshort{ue} acts as an interferer during the \acrshort{ce} phase, reducing the quality of the estimated \acrfull{csi} \cite{JOSE2009, ELIJAH2016, MARZETTA2010}.
The interference power can be several orders of magnitude greater than the signal of interest, significantly degrading the \acrfull{sinr} \cite{MARZETTA2010, LONCAR2003} and making it unadvisable to neglect the interference problem.
In various works, interference was reduced by using algorithms that strategically schedule the \acrshort{ce} phases for the different cells \cite{VU2014} or by assigning pilots in an organized way depending on the physical location of \acrshortpl{ue} \cite{MUPPIRISETTY2018}.
These methods, however, often come with a significant signaling overhead and the necessity of accurate synchronization and coordination between cells.
Conversely, methods such as \cite{WEN2015} try to estimate the channel statistical parameters and use them for computing a Bayesian estimator capable of significantly reducing the impact of \acrshort{pc}.
The authors of \cite{YIN2013} analyzed the potential performance of a Bayesian estimator used under scenarios with interference, observing that the achievable performance is strongly related to the amount of overlap between the \acrfullpl{aoa} of incoming signals.
A good estimation is possible under a large enough number of antennas if the interfering signal is spatially separated from the signal of interest, leading to orthogonal \acrfullpl{ccm}.
Several \acrfull{dl}-based methods, and in particular model-based approaches, have demonstrated excellent performance in the \acrshort{ce} task \cite{WEN2015, HUANG2018, HE2018, YE2018, SOLTANI2019}.
Our proposed solution follows the recent work on \acrshort{vae}-based channel estimators in \cite{BAUR2022, BAUR2024} and extends the analysis to the multi-cell setup.

This paper has the following contribution.
We extend the approach from \cite{BAUR2024} by considering \acrshort{pc}.
In particular, this requires adapting the original model-specific loss function to account for the new multi-cell setup and adjusting the relevant variables accordingly.
These changes are also reflected in the architecture of the \acrshort{ml} model.
Additionally, we present an analysis that considers the achievable performance depending on the amount of overlap between \acrshortpl{aoa} of incoming signals, focusing particularly on the proposed \acrshort{vae}-based method.
The proposed approach works under regular system operation without requiring coordination between cells.
Moreover, it does not require access to side information regarding \acrshortpl{ue} or other measures such as \acrfull{aoa} estimation.

\section{System Model}
\label{sec:system_and_channel_model}
We consider a \acrfull{tdd} multi-user \acrfull{simo} setup with \( L \) cells, each having a single \acrfull{bs} equipped with \( M \) antennas, with single-antenna \acrshortpl{ue} using uplink training pilot signals of length \( T_{\text{tr}} \).
During \acrshort{ce}, the \( k \)-th \acrshort{ue} in the \( l \)-th cell sends a pilot sequence \( \bm{\psi}_{k,l} \in \mathbb{C}^{T_{\text{tr}}} \) to its own \acrshort{bs}, received by the \acrshort{bs} in the cell of interest via the channel \( \bm{h}_{k,l} \in \mathbb{C}^M \).
The pilots and the channels of the \( K \) \acrshortpl{ue} in the \( l \)-th cell can be arranged into matrices as, respectively, \( \bm{\Psi}_l = \begin{bmatrix} \bm{\psi}_{1,l} & \bm{\psi}_{2,l} & \cdots & \bm{\psi}_{K,l} \end{bmatrix} \in \mathbb{C}^{T_{\text{tr}} \times K} \) and \( \bm{H}_l = \begin{bmatrix} \bm{h}_{l, 1} & \bm{h}_{l, 2} & \cdots & \bm{h}_{l, K} \end{bmatrix} \in \mathbb{C}^{M \times K} \).
When considering only the \( l \)-th cell, noiseless pilots are received at the \acrshort{bs} of the cell of interest as \( \bm{H}_l\bm{\Psi}_l\hermitian \in \mathbb{C}^{M \times T_{\text{tr}}} \).
When \( L \) cells transmit pilots at the same time and under the presence of noise, the received pilot signal at the \acrshort{bs} of the cell of interest becomes \( \bm{Y} = \sum_{l = 1}^{L}\bm{H}_l \bm{\Psi}_l\hermitian + \bm{N} \in \mathbb{C}^{M \times T_{\text{tr}}} \), where \( \bm{N} \) is complex \acrfull{awgn}.
We conservatively assume that the same set of pilots is reused among different cells, i.e., \( \bm{\Psi}_l = \bm{\Psi} \) for all \( l \).
If we consider the pilots in \( \bm{\Psi} \) to be pair-wise orthogonal, thereby assuming \( T_{\text{tr}} \geq K \) and thus absence of intracell interference, an observation of the channel of the \( k \)-th \acrshort{ue} in the \( l \)-th cell can be obtained as \( \bm{y}_{k,l} = \bm{Y}\bm{\psi}_{k,l} \).

We can rewrite \( \bm{y}_{k,l} \) omitting the subscripts, where \( k \) can be omitted to indicate a generic \acrshort{ue} in the cell of interest, and \( l \) is omitted as we are always considering the cell of interest where \( l = 1 \).
We thus write \( \bm{y}_{k,l} \) as
\begin{equation}
    \bm{y} = \bm{h}_1 + \sum_{l=2}^{L} \bm{h}_l + \bm{n} \in \mathbb{C}^M \, ,
    \label{eq:observation}
\end{equation}
where \( \bm{h}_l \) is the channel realization for the \acrshort{ue} in the \( l \)-th cell, with the subscript \( k \) being omitted once more, similarly to \( \bm{y}_{k,l} \), to consider a generic \acrshort{ue} in the cell. The variable \( \bm{n} \sim \Normal_{\mathbb{C}}(\bm{0}, \sigma^2\Identity_M) \) represents the \acrshort{awgn}.
The objective of \acrshort{ce} is estimating the channel of the \acrshort{ue} in the cell of interest, based on the observation \( \bm{y} \).

\section{Channel Estimation With VAE}

\subsection{MMSE Channel Estimation in Multi-Cell}
\label{subsec:mmse-channel-estimation-in-multi-cell}
The \acrfull{lmmse} estimation of a channel \( \bm{h}_1 \), given its observation \( \bm{y} \), is obtained using the expression
\begin{equation}
    \bm{\Hat{h}}_1(\bm{y}) = \bm{\mu}_1 + \bm{C}_{1,\bm{y}} \bm{C}_{\bm{y}}^{-1}(\bm{y} - \bm{\mu}_{\bm{y}}) \, ,
    \label{eq:vae-cme-step}
\end{equation}
where \( \bm{\mu}_{\bm{y}} \) and \( \bm{C}_{\bm{y}} \) are the observation's mean and covariance, \( \bm{\mu}_1 \) is the channel of interest mean, and \( \bm{C}_{1,\bm{y}} \) is the covariance between the channel of interest and the observation  \cite[Ch. 12]{KAY1993}.
Under a multi-cell scenario where \eqref{eq:observation} holds and different cells' channel realizations \( \bm{h}_l \) are uncorrelated with each other, the expression \eqref{eq:vae-cme-step} can be rewritten as
\begin{equation}
    \bm{\Hat{h}}_1(\bm{y}) = \bm{\mu}_1 + \bm{C}_1 \bm{C}_{\bm{y}}^{-1}(\bm{y} - \bm{\mu}_{\bm{y}}) \, ,
    \label{eq:vae-cme}
\end{equation}
where \( \bm{\bm{\mu}}_{\bm{y}} = \bm{\bm{\mu}}_1 + \sum_{l=2}^L \bm{\bm{\mu}}_l \) and \( \bm{C}_{\bm{y}} = \bm{C}_1 + \sum_{l=2}^L \bm{C}_l + \sigma^2\Identity \).
Assuming that channel realizations can be described as Gaussian, the \acrshort{lmmse} estimator in \eqref{eq:vae-cme} is \acrfull{mse}-optimal, and corresponds to the \acrshort{cme} \( \Expectation\left[ \bm{h}_1 \cond \bm{y} \right] \).
It is evident that we need the first and second statistical moments of channel realizations in the different cells for computing \eqref{eq:vae-cme}.

\subsection{VAE for Multi-Cell Channel Estimation}
\begin{figure}[t]
    \centering
    \scalebox{0.85}{\includegraphics[]{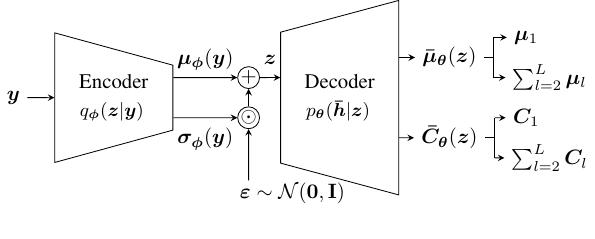}}
    \vspace{-10pt}
    \caption{\acrshort{vae} architecture for the multi-cell \acrshort{ce} problem.}
    \label{fig:multi-cell-vae}
\end{figure}

When used as a generative model, a \acrshort{vae} can generate samples from an unknown distribution \( p(\bm{h}) \), given a sample \( \bm{h} \) from the same distribution or its observation \( \bm{y} \).
This is done by approximating the unknown distribution as \acrshort{cg} \cite{KINGMA2014}.
In the \acrshort{ce} problem, we are not interested in generating new samples from \( p(\bm{h}) \); instead, we need access to the \acrshort{cg} channel distributions' means and covariances.
In particular, as mentioned in Section \ref{subsec:mmse-channel-estimation-in-multi-cell}, for the multi-cell case, we need the first two statistical moments for the cell of interest and interfering cells; thus, we produce them as the output of the \acrshort{vae}'s decoder as depicted in Fig. \ref{fig:multi-cell-vae}.
It is important to note that the unknown true channel distributions can have any form and do not need to be Gaussian to be approximated as \acrshort{cg} by the \acrshort{vae}.

Specifically, we consider a joint distribution \( p(\bm{\Bar{h}}) \), where
\begin{equation*}
    \bm{\Bar{h}} = \begin{bmatrix}
        \bm{h}_{1} \\
        \sum_{l = 2}^{L}\bm{h}_{l}
    \end{bmatrix}
\end{equation*}
is the joint variable characterized by its mean and covariance
\begin{equation*}
    \bm{\Bar{\mu}}_{\bm{\theta}}(\bm{z}) = \begin{bmatrix}
        \bm{\mu}_{1} \\
        \sum_{l = 2}^{L}\bm{\mu}_{l}
    \end{bmatrix}
    \quad \text{and} \quad
    \bm{\Bar{C}}_{\bm{\theta}}(\bm{z}) = \begin{bmatrix}
        \bm{C}_{1} & \bm{0}  \\
        \bm{0}            & \sum_{l = 2}^{L}\bm{C}_{l}
    \end{bmatrix} \,.
\end{equation*}
The joint distribution \( p(\bm{\Bar{h}}) \) remains \acrshort{cg}, as it is composed of the \acrshort{cg} channel of interest distribution and the sum of independent \acrshort{cg} interference distributions from the different interfering cells.
In addition, we enforce the covariance matrix \( \bm{\Bar{C}}_{\bm{\theta}}(\bm{z}) \) to have a block-diagonal structure, leading to an absence of correlation between the channel of interest and interference and, thus, enabling the use of \eqref{eq:vae-cme} instead of \eqref{eq:vae-cme-step}.

The parameters \( \bm{\theta }\), corresponding to the decoder \acrfull{nn} weights, allow parameterizing the \acrshort{cg} distribution \( p_{\bm{\theta}}(\bm{\Bar{h}} \cond \bm{z}) = \Normal(\bm{\Bar{\mu}}_{\bm{\theta}}(\bm{z}),\bm{\Bar{C}}_{\bm{\theta}}(\bm{z})) \).
The conditioning variable \( \bm{z} \) is obtained by sampling from the distribution \( q_{\bm{\phi}}(\bm{z} \cond \bm{y}) = \Normal(\bm{\mu}_{\bm{\phi}}(\bm{y}), \diag{(\bm{\sigma}_{\bm{\phi}}(\bm{y}))}) \) which is parameterized by the \acrshort{vae}'s encoder weights \( \bm{\phi} \).

The described \acrshort{vae} is trained by maximizing the likelihood
\begin{equation*}
    \log p_{\bm{\theta}}(\bm{\Bar{h}}) = \mathcal{L}_{\bm{\theta}, \bm{\phi}}(\bm{\Bar{h}}) + \kldiv{q_{\bm{\phi}}(\bm{z} \cond \bm{y})}{p(\bm{z})} \, ,
\end{equation*}
where \( D_{\text{KL}} \) is the \acrfull{kl} divergence between two distributions.
Instead of maximizing the likelihood, we can maximize the \acrfull{elbo} term:
\begin{equation}
    \mathcal{L}_{\bm{\theta}, \bm{\phi}}(\bm{\Bar{h}}) = \Expectation_{q_{\bm{\phi}}}\left[ \log p_{\bm{\theta}}(\bm{\Bar{h}} | \bm{z}) \right] - \kldiv{q_{\bm{\phi}}(\bm{z} | \bm{y})}{p(\bm{z})} \, ,
    \label{eq:elbo}
\end{equation}
which serves as the loss function during training.
The \acrshort{elbo} differentiation w.r.t. the \acrshort{nn}'s parameters, allowing the model's training, is made possible through the well-known \textit{reparameterization trick} \cite{KINGMA2014}.
The Gaussian nature of the involved distributions enables deriving analytic expressions for the \acrshort{elbo} in \eqref{eq:elbo}, as extensively described in \cite{BAUR2024}.

\subsection{Implementation Details}
The architecture we use for the encoder and decoder \acrshortpl{nn} is almost identical to the one used in \cite{BAUR2024}, with the only difference being the doubled sizes of the hidden-space variable and the decoder layers, allowing them to represent the additional interference statistical parameters.
As part of the encoder and decoder, we transform the model's input and output with the \acrshort{dft} and \acrfull{idft}, respectively, to help the \acrshort{nn} better exploit the sparsity of channel realizations in the angular domain.
The training of our model is performed using the Adam optimizer with a learning rate of \num{1e-4}, a batch size of \num{128}, adopting early stopping when the value of \eqref{eq:elbo} evaluated on a validation dataset does not improve for \num{100} epochs, and tuning the hyperparameters using a random search \cite{BERGSTRA2012}.

The implementation of our model is based on the \textit{\acrshort{vae}-real} variant described in \cite{BAUR2024}.
The \textit{\acrshort{vae}-real} model from \cite{BAUR2024} does not require access to ground truth data during either training or evaluation; however, our adapted model requires access, only during training, to noisy observations of the signal of interest without interference, as well as noisy observations of the interference without the signal of interest.
Such variables can be described by the joint variable \( \bm{\Bar{y}} \), defined as
\begin{equation*}
    \bm{\Bar{y}} = \begin{bmatrix}
        \bm{h}_1 + \bm{n}_1 \\
        \sum_{l=2}^L\bm{h}_l + \bm{n}_2
    \end{bmatrix} \, .
\end{equation*}
In particular, despite using  \( \bm{\Bar{y}} \) instead of \( \bm{\Bar{h}} \) in the analytic expression of \eqref{eq:elbo}, which would cause the model to learn the unwanted distribution \( p_{\bm{\theta}}(\bm{\Bar{y}}\cond\bm{z}) \), the decoder parameterizes the desired \acrshort{cg} distribution \( p_{\bm{\theta}}(\bm{\Bar{h}}\cond\bm{z}) \) using a simple workaround, similar to the one used in \cite{BAUR2024}.
We describe the zero-mean noise component in \( \bm{\Bar{y}} \) by its covariance \( \bm{\Sigma} = \diag(\left[\sigma^2_1\bm{1}\transpose \;\; \sigma^2_2\bm{1}\transpose \right]) \), with \( \sigma_1 \) and \( \sigma_2 \) being the standard deviations of \( \bm{n}_1 \) and \( \bm{n}_2 \), and \( \bm{1} \) being the all-ones vector.
Afterward, we force the decoder to obtain the distribution parameters of \( p_{\bm{\theta}}(\bm{\Bar{h}}\cond\bm{z}) \) by adding the known noise covariance \( \bm{\Sigma} \) to the covariance produced by the decoder \( \bm{\Bar{C}}_{\bm{\theta}}(\bm{z}) \) when the latter is being used as part of the analytic expression of \eqref{eq:elbo}.
The mean is left unaltered as \( p_{\bm{\theta}}(\bm{\Bar{h}}\cond\bm{z}) \) and \( p_{\bm{\theta}}(\bm{\Bar{y}}\cond\bm{z}) \) have identical mean values due to the zero-mean \acrshort{awgn}.

\section{Related Estimators}
We compare our estimators' performance with related ones.
The least-squares estimator, denoted with ``\acrshort{ls},'' is defined as 
\begin{equation*}
    \bm{\hat{h}}_{1,\text{\acrshort{ls}}}(\bm{y}) = \bm{y} \, ,
\end{equation*}
with \( \bm{y} \) being the observation from \eqref{eq:observation}.
Please note that, considering the received signal \( \bm{Y} \) and the pilot matrix \( \bm{\Psi} \) as defined in Section \ref{sec:system_and_channel_model},  the \acrshort{ls} estimator can be obtained as \( \bm{Y} \bm{\Psi} \left( \bm{\Psi}\hermitian \bm{\Psi} \right)^{-1} \), where we consider only a single column corresponding to the observation we are interested in.

The sample covariance estimator, denoted with ``scov,'' is built considering the sample covariance matrices for the \acrshortpl{ue} in each cell, computed as \( \bm{\Hat{C}}_l = 1/T_{\text{r}} \sum_{i=1}^{T_{\text{r}}} \bm{h}_{l,i}\bm{h}_{l,i}\hermitian \), together with their sample means \( \bm{\Hat{\mu}}_l = 1/T_{\text{r}} \sum_{i=1}^{T_{\text{r}}} \bm{h}_{l,i} \), where \( T_{\text{r}} \) channel realizations \( \bm{h}_{l,i} \) are considered.
Please note that computing the sample means and covariances requires access to ground truth channel realizations from each cell, which is not required by the proposed method.
We use the computed statistical parameters to define the estimator as
\begin{equation*}
    \bm{\Hat{h}}_{1, \text{scov}}(\bm{y}) = \bm{\hat{\mu}}_1 + \bm{\hat{C}}_1 \bm{\hat{C}}_{\bm{y}}^{-1}(\bm{y} - \bm{\hat{\mu}}_{\bm{y}}) \, ,
\end{equation*}
where \( \bm{\hat{\mu}}_{\bm{y}} = \sum_{l=1}^L \bm{\hat{\mu}}_l \) and \( \bm{\hat{C}}_{\bm{y}} = \sum_{l=1}^L \bm{\hat{C}}_l + \sigma^2\Identity \).

For channel models where access to the ground-truth \acrshortpl{ccm} of channel realizations is possible, we define an estimator which uses such \acrshortpl{ccm} in \eqref{eq:vae-cme} and denote it with ``genie-cov.''

We compare our method also with other \acrshort{vae}-based approaches.
We train one single-cell \acrshort{vae} as in \cite{BAUR2024}, for retrieving the statistical parameters for the cell of interest only, i.e., \( \bm{\mu}_1 \) and \( \bm{C}_1 \), but using the observations \( \bm{y} \), corrupted by interference, as input to the encoder.
We define a ``\acrshort{vae}, ignore-interference'' estimator as
\begin{equation}
    \bm{\hat{h}}_{1, \text{\acrshort{vae}-ignore}}(\bm{y}) = \bm{\mu}_1 + \bm{C}_1 (\bm{C}_1 + \sigma^2\Identity)^{-1}(\bm{y} - \bm{\mu}_1) \, ,
    \label{eq:vae-ignore-int-estimator}
\end{equation}
where the interference mean and covariance terms are completely ignored.
Another estimator, called ``\acrshort{vae}, \acrshort{awgn}-interference,'' is obtained by replacing the interference distribution covariance with a scaled identity matrix \( \alpha^2\Identity \), where the parameter \( \alpha \) is set such that the trace of the scaled identity equals the trace of the interference sample covariance matrix.
The estimator is then obtained as
\begin{equation}
    \bm{\hat{h}}_{1, \text{\acrshort{vae}-\acrshort{awgn}}} (\bm{y}) = \bm{\mu}_1 + \bm{C}_1 (\bm{C}_1 + \alpha^2\Identity + \sigma^2\Identity)^{-1}(\bm{y} - \bm{\mu}_1) \, .
    \label{eq:vae-awgn-int-estimator}
\end{equation}
The last estimator based on the single-cell \acrshort{vae} is the ``\acrshort{vae}-scov,'' which approximates the interference mean and covariance with sample mean and covariance of the training interference channels.
This estimator is thus obtained as
\begin{equation}
    \bm{\hat{h}}_{1, \text{\acrshort{vae}-scov}} (\bm{y}) = \bm{\mu}_1 + \bm{C}_1 \bm{\Tilde{C}}_{\bm{y}}^{-1}(\bm{y} - \bm{\Tilde{\mu}}_{\bm{y}}) \, ,
    \label{eq:vae-scov-estimator}
\end{equation}
where \(  \bm{\Tilde{C}}_{\bm{y}} = \bm{C}_1 + \sum_{l=2}^L \bm{\hat{C}}_l + \sigma^2\Identity \) and \( \bm{\Tilde{\mu}}_{\bm{y}} = \bm{\mu}_1 + \sum_{l=2}^L \bm{\hat{\mu}}_l \).

Finally, we introduce an alternative multi-cell \acrshort{vae}-based estimator, which we call ``\acrshort{vae}, \textit{genie}.''
This estimator is identical to the proposed one, except for using noiseless observations during training and as input to the encoder.
Although this estimator is purely theoretical, as it requires knowledge of the true \acrshort{csi} even at evaluation time, it remains useful by demonstrating the estimator's potential performance achievable when the \acrshort{vae} produces sufficiently accurate mean and covariance distribution parameters.

\section{Simulation Results}
\label{sec:simulation_results}

We create datasets consisting of \( T_{\text{r}} = \num{100000} \) training, \( T_{\text{v}} = \num{10000} \) validation, and \( T_{\text{e}} = \num{10000} \) test samples.
Data is normalized to have a per-antenna signal power of \num{1}, where we only consider the channel of interest as the signal.
We define the \acrfull{snr} as \( 1/\sigma^2\), and we evaluate our results comparing the \acrfull{nmse} values computed as \( \frac{1}{M T_{\text{e}}} \sum_{i=1}^{T_{\text{e}}} \lVert \bm{h}_{1,i} - \bm{\hat{h}}_1(\bm{y}_i) \rVert^2 \), where \( \bm{h}_{1,i} \) and \( \bm{y}_i \) are the \( i \)-th channel realization and observation, respectively, for the channel of interest in the test dataset.
Our training observations have \acrshortpl{snr} ranging from \SI{-16}{\deci\bel} to \SI{36}{\deci\bel}.

We use \acrshortpl{ccm} \( \bm{C}_{\bm{\delta}} \) constructed according to the \acrfull{3gpp} specification for the \acrfull{uma} scenario \cite{3GPPTECHREPORT}.
This model generates \acrshortpl{ccm} \( \bm{C}_{\bm{\delta}} = \int_{-\pi}^{\pi} g(\vartheta; \bm{\delta}) \bm{a}(\vartheta) \bm{a}(\vartheta)\hermitian \dif\vartheta \), where the function \( g \) represents the power angular spectrum and is parameterized by \( \bm{\delta} \), while \( \bm{a}(\vartheta) \) is the \acrshort{ula}'s array steering vector.
The parameter \( \bm{\delta} \) is a random variable that describes the propagation clusters' \acrshortpl{aoa} following a predefined distribution \( p(\bm{\delta}) \).
Channel realizations for the \( l \)-th cell are generated as \( \bm{h}_l \cond \bm{\delta}_l \sim \Normal_{\mathbb{C}}(\bm{0}, \bm{C}_{\bm{\delta}_l}) \).
Note that \( \bm{\delta}_l \) is different for every channel realization \cite{NEUMANN2018}.
We simulate a multi-cell setup by using different prior distributions \( p(\bm{\delta}_l) \) for the \acrshortpl{aoa} of multi-path signals of \acrshortpl{ue} in different cells.
In particular, we define a scenario with one \acrshort{ue} of interest and one interfering \acrshort{ue}, where we consider four different possible distribution pairs \(  p(\bm{\delta}_1) \) and \(  p(\bm{\delta}_2) \).
The first possibility considers both distributions as uniform over the entire angular domain, i.e., \( p(\bm{\delta}_l) = \Uniform \left( \SI{-180}{\degree}, \SI{180}{\degree} \right) \), while the remaining consider two normal distributions for \(  p(\bm{\delta}_1) \) and \(  p(\bm{\delta}_2) \), centered at \SI{45}{\degree} and \SI{-45}{\degree} respectively, and with both standard deviations chosen either as \SI{90}{\degree}, \SI{60}{\degree}, or \SI{30}{\degree}.
With the described setup, signal and interference have the same average power, and the different distributions allow having different amounts of overlap between the considered signals' \acrshortpl{aoa}.
For the \acrshort{3gpp} channel model, we consider 128 antennas at the BS.
Furthermore, please note that the access to the ground-truth \acrshortpl{ccm} enables the calculation of the ``genie-cov'' estimator.

A second channel model we adopt is obtained from the \acrshort{quadriga} channel simulator \cite{QUADRIGATECHREPORT}.
\Acrshort{quadriga} generates channel realizations as a superposition of \( P \) propagation paths as \( \bm{h}_l = \sum_{p=1}^{P} \bm{g}_p \exp{(-2\pi \imaginaryJ f_{\text{c}} \tau_p)} \), where \( \bm{g}_p \) is a complex-valued gain between the \acrshort{bs}'s and \acrshort{ue}'s antennas accounting for path attenuation, antenna radiation pattern, and polarization.
The carrier frequency is \( f_{\text{c}} \) and is set to \SI{6}{\giga\hertz}, while \( \tau_p \) is the delay of the \(  p \)-th propagation path.
We generate data for the \acrshort{uma} scenario with a bandwidth of \SI{180}{\kilo\hertz}.
A simple multi-cell pattern with two cells is considered, where the \acrshortpl{ue} are placed in \SI{120}{\degree} circular sectors with \SI{500}{\meter} radius.
The \acrshort{bs} is equipped with a ``\acrshort{3gpp}-3D'' antenna array with \num{32} antennas, while \acrshortpl{ue} use single omnidirectional antennas.
The \acrshort{bs} is at a height of \SI{25}{\meter}, the cell of interest and the interfering cell are placed at a distance of approximately \SI{866}{\meter} from the \acrshort{bs}, and at \SI{\pm60}{\degree} azimuth angles, such that interference has the same average power as the signal of interest.
We perform a normalization of channel realizations \( \bm{h}_l \) forming each observation \( \bm{y} \) w.r.t. the cumulative path gain of the channel realizations.
Such normalization preserves the proportion between the power of the channel of interest and interference in each observation.

For both channel models, we assume the \acrshort{bs} to be equipped with a \acrfull{ula} of \( M \) antennas, where \( M \) is large, such that the \acrshortpl{ccm} have a Toeplitz structure and the circulant matrix approximation holds \cite{GRAY2006}.

\begin{figure}[t]
    \centering
    \includegraphics[width=\linewidth]{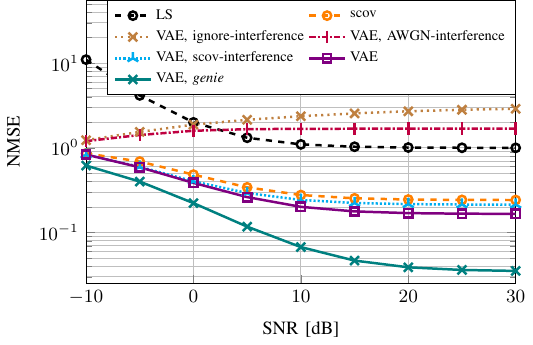}
    \caption{\acrshort{nmse} over \acrshort{snr} for \acrshort{vae} \textit{noisy} models, with \acrshort{quadriga} data and setup described in Section \ref{sec:simulation_results}.}
    \label{fig:results-quadriga}
\end{figure}

In Fig. \ref{fig:results-quadriga}, we compare the \acrshort{nmse} values for different \acrshortpl{snr} obtained with our estimator and with the related ones.
The estimators based on single-cell \acrshortpl{vae}, \eqref{eq:vae-ignore-int-estimator}, \eqref{eq:vae-awgn-int-estimator} and \eqref{eq:vae-scov-estimator}, which present an increasing level of knowledge regarding the statistical properties of interference, have decreasing \acrshort{nmse} values, demonstrating that neither we can neglect noise, nor we can approximate it as \acrshort{awgn}.
The proposed estimator, referred to as ``\acrshort{vae}'' in the results, shows better performance when compared to the single-cell \acrshort{vae}-based estimators or the classical ones.
This is explained by the improved quality of the interference statistics obtained with the proposed method.
The performance of ``\acrshort{vae}, \textit{genie}'' is presented as a benchmark and represents the potential performance of the estimator \eqref{eq:vae-cme}.
This performance is not realized in practice due to the presence of noise during the model training and at evaluation time.

\begin{figure}[t]
    \centering
    \includegraphics[width=\linewidth]{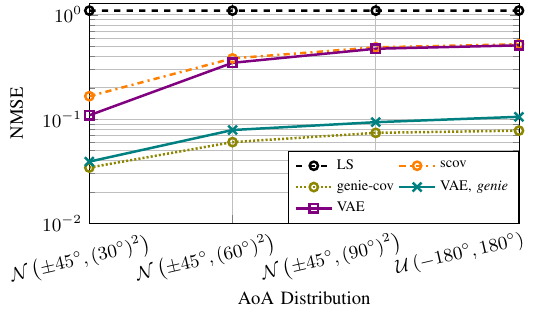}
    \caption{\acrshort{nmse} over different \acrshort{aoa} distributions, with \acrshort{3gpp} data and  \( \text{\acrshort{snr}} = \SI{10}{\deci\bel} \).}
    \label{fig:results-3gpp}
\end{figure}

Fig. \ref{fig:results-3gpp} shows the \acrshort{nmse} values for the different \acrshort{3gpp} datasets, evaluated at a fixed \acrshort{snr} level.
The \acrshort{vae}-based approach can retrieve better distribution parameters, producing better-performing estimators than the ``scov'' estimator.
This holds especially when \acrshortpl{aoa} of multi-path signals are better spatially separated.
Similarly to the results in Fig. \ref{fig:results-quadriga}, ``\acrshort{vae}, genie'' presents a theoretically achievable performance, which in this case is comparable to the one achieved by the ``genie-cov'' estimator which uses the true mean and \acrshort{ccm} of the channel realizations.


\section{Conclusion}
This work presents a \acrshort{vae}-based method for performing multi-cell \acrshort{ce}, where \acrshort{pc} and noise make the problem intricate to solve.
The proposed method extends previous work on single-cell \acrshort{vae}-based \acrshort{ce} and is capable of obtaining second-order statistical parameters for channel realizations of \acrshortpl{ue} of interest and interfering \acrshortpl{ue}, thereby estimating channels of interest with superior performance compared to other classical methods.
In future works, we want to extend the evaluation to additional relevant scenarios with different system parameters or under the availability of additional side information, further highlighting the versatility unlocked by \acrshortpl{vae} for the \acrshort{ce} task.


\end{document}